\documentclass[aps,showpacs,twocolumn]{revtex4}
\usepackage{amsmath}
\usepackage{amsfonts}
\usepackage{amssymb}
\usepackage{graphicx}

\usepackage{indentfirst}

\begin{document}

%%%%%%%%%%%%%%%%%%%%%%%%%%%%%%%%%%%%%%%%%%%%%%%%%%
%%										Header										%%
%%%%%%%%%%%%%%%%%%%%%%%%%%%%%%%%%%%%%%%%%%%%%%%%%%

\title{From recollisions to the knee: A road map for double ionization in intense laser fields}

\author{F. Mauger$^{1,2}$, C. Chandre$^1$, T. Uzer$^3$}
\affiliation{$^1$ Centre de Physique Th\'eorique, CNRS -- Aix-Marseille Universit\'es, Campus de Luminy, case 907, F-13288 Marseille cedex 09, France \\ $^2$ Ecole Centrale de Marseille, Technop\^ole de Ch\^ateau-Gombert, 38 rue Fr\'ed\'eric Joliot Curie
F-13451 Marseille Cedex 20, France
\\ $^3$ School of Physics, Georgia Institute of Technology, Atlanta, GA 30332-0430, USA}
\date{\today}

\begin{abstract}
We examine the nature and statistical properties of electron-electron collisions in the recollision process in a strong laser field.  The separation of the double ionization yield into sequential and nonsequential components leads to a bell-shaped curve for the nonsequential probability and a monotonically rising one for the sequential process. We identify key features of the nonsequential process and connect our findings in a simplified model which reproduces the knee shape for the probability of double ionization with laser intensity and associated trends.
\end{abstract}
\maketitle

%%%%%%%%%%%%%%%%%%%%%%%%%%%%%%%%%%%%%%%%%%%%%%%%%%
%%										Article										%%
%%%%%%%%%%%%%%%%%%%%%%%%%%%%%%%%%%%%%%%%%%%%%%%%%%

Multiple ionization by intense laser pulses often takes place as an uncorrelated sequence of single ionizations--often, that is, but not always: In fact, it has furnished one most striking surprises in physics of recent years~\cite{Beck08} when early experiments on double ionization~\cite{Fitt92, Kond93, Walk94, Corn00} clearly showed the unexpected existence -- and even dominance-- of another channel --the so-called non-sequential double ionization~(NSDI) channel--which cannot be regarded as two uncorrelated single-ionization events. What was even more surprising was the prominence of this channel: The by-now famous ``knee'' shape indicates that at some intensities, the yield from this process exceeds the contribution from the sequential channel by many orders of magnitude~\cite{Fitt92, Kond93, Walk94, Corn00}.  Knees have been observed in the yields of multiply charged ions~\cite{Laro98}, too, suggesting the significance of nonsequential channels there as well. 

Today NSDI is regarded as one of the most dramatic manifestations of electron-electron correlation in nature. 
Despite two decades of intensive research, it precise quantal mechanism~\cite{Beck96, Feue01, Lein00, Colg04, Barn03} remains an intriguing subject.
However, the general outline of the process, known as ``recollision''~\cite{Cork93, Scha93}, are clear enough~\cite{Ivan05}: The laser ionizes an electron, which picks up energy from the laser field and, upon reversal of the field, is hurled back at the core where it dislodges a second electron~\cite{Cork93, Scha93}. What is less clear is the nature of these collisions: How efficient are they, how much energy is shared during the collision, and when do they lead to double ionization? After all, not every recollision leads to double ionization, or does so right away.  These questions have a direct bearing on the NSDI probability, and of course, the resulting knee shape.

In this Letter, we examine the collision process in detail and answer these questions using classical mechanics~\cite{Sach01, Ho05-1, Ho05-2, Panf02, Maug09}, which is increasingly widely recognized as the tool of choice for ionization phenomena sufficiently high above threshold, since, remarkably, classical and quantum-mechanical collision results are virtually indistinguishable in that regime~\cite{Beck08}. This agreement is ascribed to the dominant role of correlation~\cite{Ho05-1}. After identifying characteristic features needed for double ionization, we separate the knee into two contributions which peak at differing intensities: A bell-shaped curve for the nonsequential process and a monotonically rising curve for the sequential component. As an ultimate test of our investigation, we connect the bare essentials of the collision process through a mapping which transforms initial conditions to ionization probabilities. This mapping yields the hallmark bell shape for NSDI. However, this does not exhaust the information contained in the mapping: Remarkably--and to quote just one example-- the fall-off of the NSDI probability with intensity turns out to be not solely due to the depletion of the initial sample by sequential double ionization (SDI) but also due to the decreasing efficacy of electron-electron collisions with increasing collision energy, a fact that comes out directly from our collisional model, as we will show below.

In order to investigate these issues, we consider the following Hamiltonian system~\cite{Ho05-1} describing a one--dimensional Helium atom using soft Coulomb potentials driven by a linearly polarized laser field of amplitude $E_0$ and frequency $\omega$~:
\begin{eqnarray} \label{eq:Hamiltonian}
   H \left( x, y, p_{x}, p_{y}, t \right) = \frac{p_{x}^{2}}{2} + \frac{p_{y}^{2}}{2} 
      + \frac{1}{\sqrt{ \left( x-y \right)^{2} + 1}} \nonumber \\
      -\frac{2}{\sqrt{x^{2}+1}} - \frac{2}{\sqrt{y^{2}+1}} + (x+y) E_{0} \sin  \omega t,
\end{eqnarray}
where~$x$ and~$y$ denote the position of each electron, and~$p_{x}$ and~$p_{y}$ their~(canonically) conjugate momenta.

Without the laser field~($E_{0}=0$), typical trajectories associated with Hamiltonian~(\ref{eq:Hamiltonian}) are composed of one electron close to the nucleus (the ``inner'' electron) and one electron further away (the ``outer'' electron)~\cite{Maug09}. 
 When the laser is turned on, the outer electron quickly ionizes while the inner one experiences a competition between the laser excitation and the Coulomb interaction with the nucleus. Its effective Hamiltonian is given by~\cite{Maug09}:
\begin{equation} \label{eq:InnerHamilt}
   H_{\rm in} \left( y , p_{y},t \right) = \frac{p_{y}^{2}}{2} - \frac{2}{\sqrt{y^{2}+1}} 
      + y E_{0} \sin \omega t.
\end{equation}
Typical trajectories of Hamiltonian~$H_{\rm in}$ display two distinct behaviors~: Close to the nucleus, the Coulomb interaction is strong enough to keep the electron attached to the core (therefore referred to the ``bound'' region in what follows). In contrast, further away from the nucleus~(``unbound'' region), the laser field prevails and sweeps the electron away. The separation between the two regions is very sharp which makes it easy to compute numerically the size of the bound region as a function of the laser field intensity~\cite{Maug09}. As a result, at the beginning of the laser excitation, if both electrons are in the unbound region, one can reasonably expect sequential double ionization, while if at least one electron is in the bound region, a recollision~(with the outer electron) is needed to raise it to the unbound region in order to ionize. When both electrons ionize at~(about) the same time, the process is usually labeled as nonsequential double ionization (NSDI),  whereas double ionizations with a large delay between ejections are sequential double ionizations (SDI). However, these definitions fail to take into account the correlated nature of multiple ionization processes. For instance, a recollision may put the inner electron into an almost-bound state which then takes a long time~(sometimes more than one laser cycle) to ionize. With the previous definition, these so-called ``recollision excitation with subsequent ionization'' (RESI)~\cite{Rude04, Feue01} events -- by no means rare -- would be labeled as SDI whereas they clearly correspond to a correlated process in the same way as NSDI. Rather than decomposing the double ionization yields into SDI and NSDI contributions, here we consider the decomposition into correlated double ionization~(CDI), where at least one recollision is needed for double ionization, and uncorrelated double ionization (UDI) where no recollision is needed. For a good approximation of the UDI contribution to the double ionization yield we use the distance between the inner electron to the nucleus, obtained from Hamiltonian~(\ref{eq:InnerHamilt}). This distance is best expressed in terms of its energy  
$$
H_0 \left(y, p_{y} \right)=\frac{p_{y}^{2}}{2} - \frac{2}{\sqrt{y^{2}+1}}. 
$$ 
The smaller this energy is, the closer to the nucleus the electron is. Since $H_0$ is integrable, there exists (at least locally) a canonical transformation which maps this Hamiltonian into action and angle variables so that it only depends on the actions. The action of the inner electron is defined by $A=\oint p_y dy/(2\pi)$. We denote by~$A_{m}\left(E_{0}\right)$ the action of the outermost invariant torus of Hamiltonian~(\ref{eq:InnerHamilt}). A good approximation of the unbound region is given by $\mathcal{D}\left(E_{0} \right) = \left\{\left(y,p_{y}\right)\mbox{ s.t. }A(y,p_{y})>A_m(E_0)\right\}$. The UDI probability is given by the proportion of the ground state energy surface where both electrons belong to $\mathcal{D} \left( E_{0} \right)$. 

\begin{figure}[htb]
	\centering
		\includegraphics[width = 8cm]{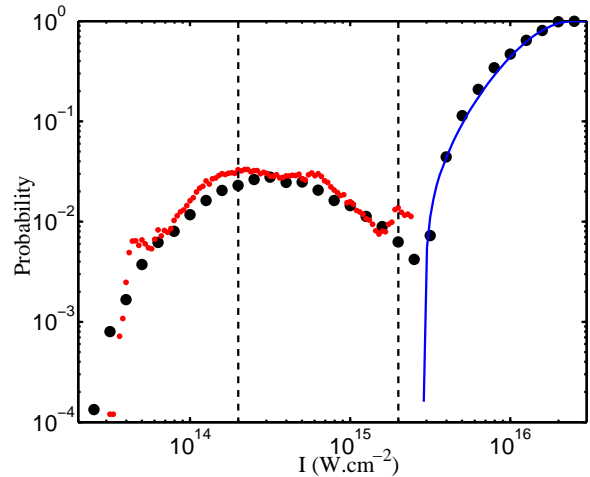}
	\caption{\label{fig:knee}
	Double ionization probability for Hamiltonian~(\ref{eq:Hamiltonian})~(big black circles) and the mapping~(\ref{eq:Mapping})~(small red circles) as a function of the laser intensity~$I$ for a wavelength of~$780$ nm. We also indicate the expected UDI probability obtained as described in the text~(continuous blue curve). Vertical dashed curves refer to intensities where phase portraits are displayed in Fig.~\ref{fig:Mapping}.}
\end{figure}

In Fig.~\ref{fig:knee}, we display the UDI component predicted by this model (continuous blue curve) which is in very good agreement with the double ionization probability obtained by integrating the full Hamiltonian~(\ref{eq:Hamiltonian}) in the high-intensity regime.

\paragraph*{Statistical analysis of recollisions.}

From the previous definition of UDI, the CDI probability is obtained by subtracting the expected UDI from the double ionization yield. 

A quick inspection of Fig.~\ref{fig:knee} reveals a bell-shaped curve for the resulting CDI component. A rather intuitive mechanism to explain the decreasing part of this bell shape is a conversion from CDI trajectories into UDI, when the laser field becomes stronger. However, in Fig.~\ref{fig:knee}, we notice a local decrease of the total yield (also observed with quantal computations~\cite{Lapp98}) which is larger by several orders than the increase of UDI. We notice that this incompatibility is readily observed in Fig.~1 of Ref.~\cite{Lapp98}. As we shall see below, the decrease of CDI is mainly due to the decrease of recollision efficiency with the laser intensity.
In order to identify the key features responsible for the decrease of CDI, and thus for the knee shape, we first collect statistical data from the trajectories associated with Hamiltonian~(\ref{eq:Hamiltonian}) at recollision times. In particular, we record the number of recollisions, the momentum of the outer electron at recollision and the energy (or rather the action) exchanged during non-ionizing recollisions.

\begin{figure}[htb]
	\centering
		\includegraphics[width = \linewidth]{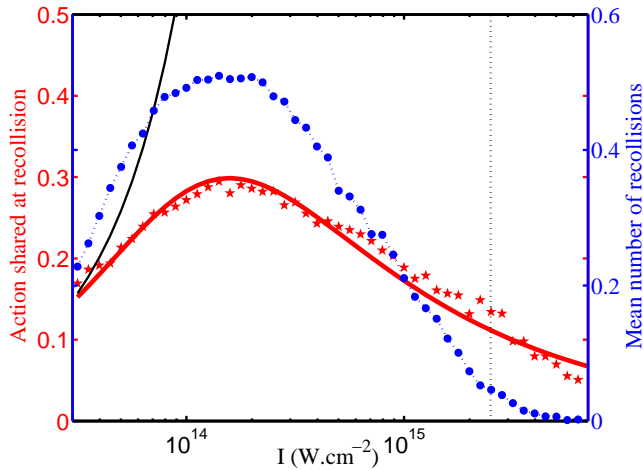}
	\caption{\label{fig:EnergySharing}
	Standard deviation~(red stars, left hand y-scale) of shared action during recollision for the full Hamiltonian~(\ref{eq:Hamiltonian}) as a function of the laser intensity $I$. An approximation of the standard deviations~(red continuous curve) is given by Eq.~(\ref{eq:ExchangedEnergyFit}). The thin black continuous line corresponds to the maximum recollision energy $\mathcal{E}_{\max}$~\cite{Cork93, Scha93} (expressed in terms of actions) and the vertical dotted line indicates the intensity after which we stop the mapping. The dashed-dotted curve (blue line, right-hand scale) shows the mean number of recollisions computed from all analyzed trajectories. The typical CDI process requires between one (low intensity) and four (high intensity) recollisions.}
\end{figure}

It is well-known that the maximum energy the outer electron can bring back to the core is equal to~\cite{Cork93,Band05} $\mathcal{E}_{\max}=\kappa E_{0}^{2}/\left(4\omega^{2}\right)$ where $\kappa\approx 3.17$. However, an inspection of the mean energy exchanged during recollisions shows that this amount is significantly smaller and decreases with intensity during (non-ionizing) recollisions (see Fig.~\ref{fig:EnergySharing}) at relatively high intensities. This is in agreement with the increase (with the laser intensity) of the mean number of recollisions required for double ionization. A closer examination of the curve reveals that the mean shared action (as well as the mean exchanged energy) at recollision decreases as~$1/E_{0}$. This decrease is also found rigorously using a simplified model of recollision between an inner electron in a harmonic potential recolliding with a free outer electron.

To combine the two trends of the mean shared action $\Delta A\left(E_{0}\right)$ (proportional to~$E_{0}^{2}$ at low intensities as given by ${\cal E}_{\max}$ and to~$1/E_{0}$ at higher ones) we fit it by:
\begin{equation} \label{eq:ExchangedEnergyFit}
   \Delta A \left( E_{0} \right) = \frac{\alpha E_{0}^{2}}{\beta + E_{0}^{3}},
\end{equation}
which combines the two features. The parameters~$\alpha$ and $\beta$ are computed so as to accurately reproduce the evolution of the mean exchanged energy during recollisions (see continuous lines in Fig.~\ref{fig:EnergySharing}), here~$\alpha = 3.5\times 10^{-2}$ and~$\beta = 1.5\times 10^{-4}$.

% ###############################################################################################################################
% ##   Recollision model mapping
% ###############################################################################################################################

\paragraph*{Mapping model for recollisions.}

A sequential double ionization involves an inner electron which is classically confined on the invariant tori of Hamiltonian~(\ref{eq:InnerHamilt})~\cite{Maug09}. In the neighborhood of the nucleus, the motion is harmonic with a frequency of $\sqrt{2}$, and moving away from the nucleus, the frequency decreases. We consider a simplified model for the inner electron in action-angle variables given by the integrable Hamiltonian $H_0(A)=-2+\sqrt{2}({\rm e}^{aA}-1)/a$ where $a$ can be adjusted for quantitative agreement with the frequencies obtained for Hamiltonian~(\ref{eq:InnerHamilt}) in the bound region (here, $a=-0.8$). This formula for $H_0$ comes from the observation that the frequency depends approximately linearly on the energy in the whole bound region. 

When the outer electron returns to the core, it gives a kick in action to the inner one, which jumps from one invariant torus to another (or to the unbound region), i.e., at each recollision, the inner electron experiences a kick. The action of the inner electron is constant between two recollisions.  This collision dynamics is modeled by the ``kicked'' rotator~\cite{Casa88}:
\begin{equation} \label{eq:HamiltMap}
      H_{\rm m} \left( \varphi, A,t \right)  = H_0(A) + 
         \varepsilon A \cos \varphi \ \sum_{n=1}^{N} {\delta \left( t - nT \right)},
\end{equation}
where~$H_0(A)$ is the integrable part of the Hamiltonian of the inner electron, and $T$ is the delay between two recollisions, which is assumed to be constant and equal to half a period of the laser field. We denote by $N$ the number of recollisions. The recollisions are modeled by a kick in action equal to $\varepsilon A \cos\varphi$ such that a kick might increase or decrease the action according to the respective phase between the two electrons. In addition, it is more difficult to kick the inner electron out if it is at the bottom of the well, so the kick strength is proportional to $A$. In this way, the action remains positive at all times. The maximum strength of the kick depends strongly on $E_0$ as given by Eq.~(\ref{eq:ExchangedEnergyFit}). 

\begin{figure}[htb]
	\centering
		\includegraphics[width = \linewidth]{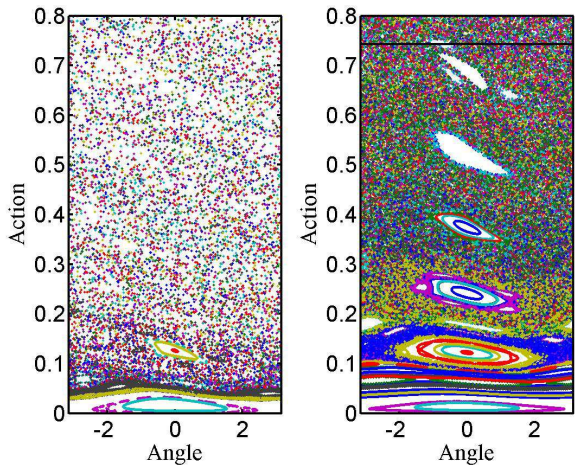}
	\caption{\label{fig:Mapping}
	Phase space portrait of some trajectories of the mapping~(\ref{eq:Mapping}) for low intensity (left panel) $I=2 \times 10^{14}\ \mbox{W}\cdot\mbox{cm}^{-2}$, and for high intensity (right panel) $I=2 \times 10^{15}\ \mbox{W}\cdot\mbox {cm}^{-2}$, represented by vertical dashed lines in Fig.~\ref{fig:knee}. In the right panel, we indicate the critical action~$A_{m}=0.74$~by a horizontal line~(whereas the critical action $A_{m}$ is $1.57$ in the left panel).}
\end{figure}

Let~$\varphi_{n}$ and~$A_{n}$ be the angle and action of a trajectory of Hamiltonian~(\ref{eq:HamiltMap}) at time~$\left(nT\right)^{-}$~(right before the $n^{th}$ kick). By integrating the trajectories between two kicks, we approximate the dynamics of Hamiltonian~(\ref{eq:HamiltMap}) by the two-dimensional symplectic map~:
\begin{equation} \label{eq:Mapping}
   \begin{array}{ccl}
      A_{n+1}       & = & A_{n}/(1- \varepsilon \sin \varphi_{n}), \\
      \varphi_{n+1} & = & \varphi_{n} + \omega_{0} \left( A_{n+1} \right) T+\varepsilon\cos \varphi_n,
   \end{array}
\end{equation}
where~$\omega_{0}\left(A\right)=\sqrt{2}{\rm e}^{a A}$ is the frequency of the inner electron.
In Fig.~\ref{fig:Mapping}, we display two phase portraits of the mapping~(\ref{eq:Mapping}) for two laser intensities~: One at low intensity ($I=2 \times 10^{14}\ \mbox{W}\cdot\mbox{cm}^{-2}$) in the range of intensity where CDI is maximum, where the phase portrait appears to be very chaotic, and one at high intensity ($I=2 \times 10^{15}\ \mbox{W}\cdot\mbox {cm}^{-2}$) where the phase portrait is more regular. In the chaotic region, the diffusion is much stronger at the maximum of CDI than for larger intensities (see Fig.~\ref{fig:Mapping}, left panel, where trajectories escape quickly from the core region, explaining that there are fewer points than in the right panel). Since the strength of the kicks decreases with the intensity at high intensities, the phase space becomes more regular. If the inner electron is inside an elliptic island (which occurs mainly at high intensities), it will not ionize regardless of the number of recollisions it undergoes. As the intensity increases, the recollisions become less effective and the map becomes integrable so fewer CDI events occur. 

% ###############################################################################################################################
% ##   Nonsequential double ionization
% ###############################################################################################################################

Through the previous mapping, we have derived a simple model for the dynamics of the inner electron initially in the bound region which experiences recollisions with the outer one. From this model, we compute ionization probabilities of the inner electron from which we deduce the probability of double ionization as a function of the laser intensity: Double ionization occurs if and only if the inner electron ionizes and we assume that the outer electron remains ionized for all times. The picture of the bound and unbound regions for the effective Hamiltonian~(\ref{eq:InnerHamilt}) of the inner electron gives a natural criterion for ionization. We recall that once the inner electron has reached an action larger than the outermost invariant torus~(with action~$A_{m}$), it is driven away from the nucleus by the laser field. Therefore, all recollisions leading to an action larger than a critical value $A_m$ (which depends on $E_0$) subsequently lead to ionization of the inner electron and, in angle-action variables, the unbound region becomes ~$\mathcal{D}\left(E_{0}\right)=\left\{\left(\varphi, A \right) \ \mbox{s.t.} \ A >A_{m}(E_0)\right\}$. 

From an initial distribution in angle-action coordinates obtained from the microcanonical distribution of inner electrons in phase space, we iterate the mapping~(\ref{eq:Mapping}) a fixed number of times for different intensities~(and thus different~$\varepsilon$). In Fig.~\ref{fig:knee}, we display the probabilities of double ionization as a function of the laser intensity.  We disregard recollisions for intensities larger than~$2.5 \times 10^{15} \ \mbox{W}\cdot\mbox
 {cm}^{-2}$. This adjustment is motivated by the weak probability of recollisions we have detected in the data analysis process~(see Fig.~\ref{fig:EnergySharing}). We notice that it qualitatively reproduces the trends observed in the double ionization yields observed using a statistical analysis of trajectories of Hamiltonian~(\ref{eq:Hamiltonian}). In particular, the asymmetry in the increase and decrease of the CDI probability is worth noting.  
 
In summary, we have identified key properties of the inelastic electron-electron collisions through the analysis of recolliding trajectories. We connected our findings to a simplified model for the dynamics of the recollisions, which amounts to a two-dimensional symplectic map in the action-angle coordinates of the inner electron. The statistical analysis of the (discrete-time) trajectories of this model results in the knee shape for the probability of double ionization versus intensity. A proper decomposition into correlated and uncorrelated double ionization yields a  bell-shape for the correlated process and a monotonic rise for the uncorrelated one.

\acknowledgements
C.C. and F.M. acknowledge financial support from the PICS program of the CNRS. This work is partially funded by NSF.

%%%%%%%%%%%%%%%%%%%%%%%%%%%%%%%%%%%%%%%%%%%%%%%%%%
%%									Bibliography								%%
%%%%%%%%%%%%%%%%%%%%%%%%%%%%%%%%%%%%%%%%%%%%%%%%%%

%%%%%%%%%%%%%%%%%%%%%%%%%%%%%%%%%%%%%%%%%%%%%%%%%%
%%%%%%%%%%%%%%%%%%%%%%%%%%%%%%%%%%%%%%%%%%%%%%%%%%


\begin{thebibliography}{23}
\expandafter\ifx\csname natexlab\endcsname\relax\def\natexlab#1{#1}\fi
\expandafter\ifx\csname bibnamefont\endcsname\relax
  \def\bibnamefont#1{#1}\fi
\expandafter\ifx\csname bibfnamefont\endcsname\relax
  \def\bibfnamefont#1{#1}\fi
\expandafter\ifx\csname citenamefont\endcsname\relax
  \def\citenamefont#1{#1}\fi
\expandafter\ifx\csname url\endcsname\relax
  \def\url#1{\texttt{#1}}\fi
\expandafter\ifx\csname urlprefix\endcsname\relax\def\urlprefix{URL }\fi
\providecommand{\bibinfo}[2]{#2}
\providecommand{\eprint}[2][]{\url{#2}}

\bibitem[{\citenamefont{Becker and Rottke}(2008)}]{Beck08}
\bibinfo{author}{\bibfnamefont{W.}~\bibnamefont{Becker}} \bibnamefont{and}
  \bibinfo{author}{\bibfnamefont{H.}~\bibnamefont{Rottke}},
  \bibinfo{journal}{Contemporary Physics} \textbf{\bibinfo{volume}{49}},
  \bibinfo{pages}{199} (\bibinfo{year}{2008}).

\bibitem[{\citenamefont{Fittinghoff et~al.}(1992)\citenamefont{Fittinghoff,
  Bolton, Chang, and Kulander}}]{Fitt92}
\bibinfo{author}{\bibfnamefont{D.~N.} \bibnamefont{Fittinghoff}} \bibnamefont{{\it et~al.}},
%  \bibinfo{author}{\bibfnamefont{P.~R.} \bibnamefont{Bolton}},
%  \bibinfo{author}{\bibfnamefont{B.}~\bibnamefont{Chang}}, \bibnamefont{and}
%  \bibinfo{author}{\bibfnamefont{K.~C.} \bibnamefont{Kulander}},
  \bibinfo{journal}{Phys.~Rev.~Lett.} \textbf{\bibinfo{volume}{69}},
  \bibinfo{pages}{2642} (\bibinfo{year}{1992}).

\bibitem[{\citenamefont{Kondo et~al.}(1993)\citenamefont{Kondo, Sagisaka,
  Tamida, Nabekawa, and Watanabe}}]{Kond93}
\bibinfo{author}{\bibfnamefont{K.}~\bibnamefont{Kondo}} \bibnamefont{{\it et~al.}},
%  \bibinfo{author}{\bibfnamefont{A.}~\bibnamefont{Sagisaka}},
%  \bibinfo{author}{\bibfnamefont{T.}~\bibnamefont{Tamida}},
%  \bibinfo{author}{\bibfnamefont{Y.}~\bibnamefont{Nabekawa}}, \bibnamefont{and}
%  \bibinfo{author}{\bibfnamefont{S.}~\bibnamefont{Watanabe}},
  \bibinfo{journal}{Phys.~Rev.~A} \textbf{\bibinfo{volume}{48}},
  \bibinfo{pages}{R2531} (\bibinfo{year}{1993}).

\bibitem[{\citenamefont{Walker et~al.}(1994)\citenamefont{Walker, Sheehy,
  DiMauro, Agostini, Schafer, and Kulander}}]{Walk94}
\bibinfo{author}{\bibfnamefont{B.}~\bibnamefont{Walker}} \bibnamefont{{\it et~al.}},
%  \bibinfo{author}{\bibfnamefont{B.}~\bibnamefont{Sheehy}},
%  \bibinfo{author}{\bibfnamefont{L.~F.} \bibnamefont{DiMauro}},
%  \bibinfo{author}{\bibfnamefont{P.}~\bibnamefont{Agostini}},
%  \bibinfo{author}{\bibfnamefont{K.~J.} \bibnamefont{Schafer}},
%  \bibnamefont{and} \bibinfo{author}{\bibfnamefont{K.~C.}
%  \bibnamefont{Kulander}}, 
  \bibinfo{journal}{Phys.~Rev.~Lett.}
  \textbf{\bibinfo{volume}{73}}, \bibinfo{pages}{1227} (\bibinfo{year}{1994}).

\bibitem[{\citenamefont{Cornaggia and Hering}(2000)}]{Corn00}
\bibinfo{author}{\bibfnamefont{C.}~\bibnamefont{Cornaggia}} \bibnamefont{and}
  \bibinfo{author}{\bibfnamefont{P.}~\bibnamefont{Hering}},
  \bibinfo{journal}{Phys.~Rev.~A} \textbf{\bibinfo{volume}{62}},
  \bibinfo{pages}{023403} (\bibinfo{year}{2000}).

\bibitem[{\citenamefont{Larochelle et~al.}(1998)\citenamefont{Larochelle,
  Talebpoury, and Chin}}]{Laro98}
\bibinfo{author}{\bibfnamefont{S.}~\bibnamefont{Larochelle}},
  \bibinfo{author}{\bibfnamefont{A.}~\bibnamefont{Talebpour}},
  \bibnamefont{and} \bibinfo{author}{\bibfnamefont{S.~L.} \bibnamefont{Chin}},
  \bibinfo{journal}{J.~Phys.~B.} \textbf{\bibinfo{volume}{31}},
  \bibinfo{pages}{1201} (\bibinfo{year}{1998}).

% #####

\bibitem[{\citenamefont{Becker et~al.}(1996)\citenamefont{Becker and
  Faisal}}]{Beck96}
\bibinfo{author}{\bibfnamefont{A.}~\bibnamefont{Becker}}
  \bibnamefont{and} \bibinfo{author}{\bibfnamefont{F.~H.~M.} \bibnamefont{Faisal}},
  \bibinfo{journal}{J.~Phys.~B.} \textbf{\bibinfo{volume}{29}},
  \bibinfo{pages}{L197} (\bibinfo{year}{1996}).
  
\bibitem[{\citenamefont{Feuerstein et~al.}(2001)\citenamefont{Feuerstein, 
   Moshammer, Fischer, Dorn, Schr\"oter, Deipenwisch, Crespo Lopez-Urrutia, 
   H\"ohr, Neumayer, Ullrich, Rottke, Trump, Wittmann, Korn and Sandner}}]{Feue01}
\bibinfo{author}{\bibfnamefont{B.}~\bibnamefont{Feuerstein}} \bibnamefont{{\it et~al.}},
  \bibinfo{journal}{Phys. Rev. Lett.} \textbf{\bibinfo{volume}{87}},
  \bibinfo{pages}{043003} (\bibinfo{year}{2001}).

% #####

\bibitem[{\citenamefont{Lein et~al.}(2000)\citenamefont{Lein, Gross and Engel}}]{Lein00}
\bibinfo{author}{\bibfnamefont{M.} \bibnamefont{Lein}},
  \bibinfo{author}{\bibfnamefont{E.~K.~U.} \bibnamefont{Gross}}
  \bibnamefont{and} \bibinfo{author}{\bibfnamefont{V.} \bibnamefont{Engel}}, 
  \bibinfo{journal}{Phys. Rev. Lett.}
  \textbf{\bibinfo{volume}{85}}, \bibinfo{pages}{4707} (\bibinfo{year}{2000}).
  
\bibitem[{\citenamefont{Colgan et~al.}(2004)\citenamefont{Colgan, Pindzola and Robicheaux}}]{Colg04}
\bibinfo{author}{\bibfnamefont{J.} \bibnamefont{Colgan}},
  \bibinfo{author}{\bibfnamefont{M.~S.} \bibnamefont{Pindzola}}
  \bibnamefont{and} \bibinfo{author}{\bibfnamefont{F.} \bibnamefont{Robicheaux}}, 
  \bibinfo{journal}{Phys. Rev. Lett.}
  \textbf{\bibinfo{volume}{93}}, \bibinfo{pages}{053201} (\bibinfo{year}{2004}).
  
\bibitem[{\citenamefont{Barna et~al.}(2000)\citenamefont{Barna and Rost}}]{Barn03}
\bibinfo{author}{\bibfnamefont{I.~F.} \bibnamefont{Barna}}
  \bibnamefont{and} \bibinfo{author}{\bibfnamefont{J.~M.} \bibnamefont{Rost}}, 
  \bibinfo{journal}{Eur.~Phys.~J.~D}
  \textbf{\bibinfo{volume}{27}}, \bibinfo{pages}{287} (\bibinfo{year}{2003}).

% #####

\bibitem[{\citenamefont{Corkum}(1993)}]{Cork93}
\bibinfo{author}{\bibfnamefont{P.~B.} \bibnamefont{Corkum}},
  \bibinfo{journal}{Phys. Rev. Lett.} \textbf{\bibinfo{volume}{71}},
  \bibinfo{pages}{1994} (\bibinfo{year}{1993}).

\bibitem[{\citenamefont{Schafer et~al.}(1993)\citenamefont{Schafer, Yang,
  DiMauro, and Kulander}}]{Scha93}
\bibinfo{author}{\bibfnamefont{K.~J.} \bibnamefont{Schafer}} \bibnamefont{{\it et~al.}},
%  \bibinfo{author}{\bibfnamefont{B.}~\bibnamefont{Yang}},
%  \bibinfo{author}{\bibfnamefont{L.~F.} \bibnamefont{DiMauro}},
%  \bibnamefont{and} \bibinfo{author}{\bibfnamefont{K.~C.}
%  \bibnamefont{Kulander}}, 
  \bibinfo{journal}{Phys. Rev. Lett.}
  \textbf{\bibinfo{volume}{70}}, \bibinfo{pages}{1599} (\bibinfo{year}{1993}).
  
\bibitem[{\citenamefont{Ivanov et~al.}(2005)\citenamefont{Ivanov, Spanner and Smirnova}}]{Ivan05}
\bibinfo{author}{\bibfnamefont{M.~Y.} \bibnamefont{Ivanov}},
  \bibinfo{author}{\bibfnamefont{M.}~\bibnamefont{Spanner}}
  \bibnamefont{and} \bibinfo{author}{\bibfnamefont{O.}~\bibnamefont{Smirnova}}, 
  \bibinfo{journal}{J. Mod. Opt.}
  \textbf{\bibinfo{volume}{52}}, \bibinfo{pages}{165} (\bibinfo{year}{2005}).

\bibitem[{\citenamefont{Sacha and Eckhardt}(2001)}]{Sach01}
\bibinfo{author}{\bibfnamefont{K.}~\bibnamefont{Sacha}} \bibnamefont{and}
  \bibinfo{author}{\bibfnamefont{B.}~\bibnamefont{Eckhardt}},
  \bibinfo{journal}{Phys. Rev. A} \textbf{\bibinfo{volume}{63}},
  \bibinfo{pages}{043414} (\bibinfo{year}{2001}).

\bibitem[{\citenamefont{Ho et~al.}(2005)\citenamefont{Ho, Panfili, Haan, and
  Eberly}}]{Ho05-1}
\bibinfo{author}{\bibfnamefont{P.~J.} \bibnamefont{Ho}} \bibnamefont{{\it et~al.}},
%  \bibinfo{author}{\bibfnamefont{R.}~\bibnamefont{Panfili}},
%  \bibinfo{author}{\bibfnamefont{S.~L.} \bibnamefont{Haan}}, \bibnamefont{and}
%  \bibinfo{author}{\bibfnamefont{J.~H.} \bibnamefont{Eberly}},
  \bibinfo{journal}{Phys. Rev. Lett.} \textbf{\bibinfo{volume}{94}},
  \bibinfo{pages}{093002} (\bibinfo{year}{2005}).

\bibitem[{\citenamefont{Ho and Eberly}(2005)}]{Ho05-2}
\bibinfo{author}{\bibfnamefont{P.~J.} \bibnamefont{Ho}} \bibnamefont{and}
  \bibinfo{author}{\bibfnamefont{J.~H.} \bibnamefont{Eberly}},
  \bibinfo{journal}{Phys. Rev. Lett.} \textbf{\bibinfo{volume}{95}},
  \bibinfo{pages}{193002} (\bibinfo{year}{2005}).

\bibitem[{\citenamefont{Panfili et~al.}(2002)\citenamefont{Panfili, Haan, and
  Eberly}}]{Panf02}
\bibinfo{author}{\bibfnamefont{R.}~\bibnamefont{Panfili}},
  \bibinfo{author}{\bibfnamefont{S.~L.} \bibnamefont{Haan}}, \bibnamefont{and}
  \bibinfo{author}{\bibfnamefont{J.~H.} \bibnamefont{Eberly}},
  \bibinfo{journal}{Phys. Rev. Lett.} \textbf{\bibinfo{volume}{89}},
  \bibinfo{pages}{113001} (\bibinfo{year}{2002}).

\bibitem[{\citenamefont{Mauger et~al.}(2009{\natexlab{a}})\citenamefont{Mauger,
  Chandre, and Uzer}}]{Maug09}
\bibinfo{author}{\bibfnamefont{F.}~\bibnamefont{Mauger}},
  \bibinfo{author}{\bibfnamefont{C.}~\bibnamefont{Chandre}}, \bibnamefont{and}
  \bibinfo{author}{\bibfnamefont{T.}~\bibnamefont{Uzer}},
  \bibinfo{journal}{Phys.~Rev.~Lett.} \textbf{\bibinfo{volume}{102}},
  \bibinfo{pages}{173002} (\bibinfo{year}{2009}{\natexlab{a}});
  \bibinfo{journal}{J.~Phys.~B.} \textbf{\bibinfo{volume}{42}},
  \bibinfo{pages}{165602}
  (\bibinfo{year}{2009}{\natexlab{b}}).

\bibitem[{\citenamefont{Rudenko et~al.}(2004)\citenamefont{Rudenko, Zrost,
  Feuerstein, de~Jesus, Schr\"oter, Moshammer, and Ullrich}}]{Rude04}
\bibinfo{author}{\bibfnamefont{A.}~\bibnamefont{Rudenko}} \bibnamefont{{\it et~al.}},
%  \bibinfo{author}{\bibfnamefont{K.}~\bibnamefont{Zrost}},
%  \bibinfo{author}{\bibfnamefont{B.}~\bibnamefont{Feuerstein}},
%  \bibinfo{author}{\bibfnamefont{V.~L.~B.} \bibnamefont{de~Jesus}},
%  \bibinfo{author}{\bibfnamefont{C.~D.} \bibnamefont{Schr\"oter}},
%  \bibinfo{author}{\bibfnamefont{R.}~\bibnamefont{Moshammer}},
%  \bibnamefont{and} \bibinfo{author}{\bibfnamefont{J.}~\bibnamefont{Ullrich}},
  \bibinfo{journal}{Phys. Rev. Lett.} \textbf{\bibinfo{volume}{93}},
  \bibinfo{pages}{253001} (\bibinfo{year}{2004}).

% #####

\bibitem[{\citenamefont{Lappas et~al.}(1996)\citenamefont{Lappas and
  van Leeuwen}}]{Lapp98}
\bibinfo{author}{\bibfnamefont{D.~G.}~\bibnamefont{Lappas}}
  \bibnamefont{and} \bibinfo{author}{\bibfnamefont{R.} \bibnamefont{van Leeuwen}},
  \bibinfo{journal}{J.~Phys.~B.} \textbf{\bibinfo{volume}{31}},
  \bibinfo{pages}{L249} (\bibinfo{year}{1998}).

% #####

\bibitem[{\citenamefont{Bandrauk et~al.}(2005)\citenamefont{Bandrauk,
  Chelkowski, and Goudreau}}]{Band05}
\bibinfo{author}{\bibfnamefont{A.~D.} \bibnamefont{Bandrauk}},
  \bibinfo{author}{\bibfnamefont{S.}~\bibnamefont{Chelkowski}},
  \bibnamefont{and} \bibinfo{author}{\bibfnamefont{S.}~\bibnamefont{Goudreau}},
  \bibinfo{journal}{Mod.~Opt.} \textbf{\bibinfo{volume}{52}},
  \bibinfo{pages}{411 } (\bibinfo{year}{2005}).

\bibitem[{\citenamefont{Casati et~al.}(1988)\citenamefont{Casati, Guarneri, and
  Shepelyansky}}]{Casa88}
\bibinfo{author}{\bibfnamefont{G.}~\bibnamefont{Casati}},
  \bibinfo{author}{\bibfnamefont{I.}~\bibnamefont{Guarneri}}, \bibnamefont{and}
  \bibinfo{author}{\bibfnamefont{D.}~\bibnamefont{Shepelyansky}},
  \bibinfo{journal}{IEEE~J.~Quant.~Elec.} \textbf{\bibinfo{volume}{24}}, \bibinfo{pages}{1420}
  (\bibinfo{year}{1988}).

\end{thebibliography}
\end{document}